\documentclass[11pt]{article}
\usepackage{cite}

\usepackage{amsmath,amsfonts,amsthm,amssymb}
\usepackage{epsfig}  

\def\ii{\'\i}
\def\ao{\~ao}

\def\cao{\c c\~ao}

\def\ii{\'\i}
\def\ao{\~ao}

\def\cao{\c c\~ao}

\def\ftoday{{\sl {Le \number\day \space\ifcase\month 
\or janvier\or f\'evrier\or mars\or avril\or mai
\or juin\or juillet\or ao\^ut\or septembre\or octobre
\or novembre \or d\'ecembre\fi\space \number\year}}}    
\def\ptoday{{\sl {\number\day \space de\space \ifcase\month 
\or janeiro\or fevereiro\or mar{\c c}o\or abril\or maio
\or junho\or julho\or agosto\or setembro\or outubro
\or novembro \or dezembro\fi\space de\space \number\year}}}    
\def\gtoday{{\sl {Den \number\day. \ifcase\month 
\or Januar\or Februar\or M\"arz\or April\or Mai
\or Juni\or Juli\or August\or September\or Oktober
\or November \or Dezember\fi\space \number\year}}}    
\def\today{{\sl {\ifcase\month
\or January\or February\or March\or April\or May
\or June\or July\or August\or September\or October
\or November \or December\fi \space\number\day,\space 
                                            \number\year}}}
\newcommand{\journal}[4]{{\em #1~}#2\,(#3)\,#4}

\newcommand{\pr}{\journal {Phys. Rev.}}

\newcommand{\jpcs}{\journal {J. Phys. Conf. Ser.}}

\newcommand{\cqg}{\journal {Class. Quantum Grav.}}

\newcommand{\np}{\journal {Nucl. Phys.}}
\newcommand{\pl}{\journal {Phys. Lett.}}

\newcommand{\jgp}{\journal {J. Geom. Phys.}}

\renewcommand{\a}{\alpha}
\renewcommand{\b}{\beta}
\newcommand{\g}{\gamma}           
\renewcommand{\d}{\delta}         
\newcommand{\e}{\varepsilon}

\newcommand{\la}{\lambda}        
\newcommand{\m}{\mu}

\newcommand{\om}{\omega}

\renewcommand{\AA}{{\cal A}}

\newcommand{\GG}{{\cal G}}

\newcommand{\LL}{{\cal L}}

\newcommand{\OO}{{\cal O}}

\newcommand{\SSS}{{\cal S}}

\newcommand{\es}{\\[3mm]}

\newcommand{\sla}{\raise.15ex\hbox{$/$}\kern -.57em} 
\newcommand{\Sla}{\raise.15ex\hbox{$/$}\kern -.70em}

\def\Lp{\displaystyle{\biggl(}}
\def\Rp{\displaystyle{\biggr)}}

\newcommand{\lp}{\left(}\newcommand{\rp}{\right)}

\newcommand{\lac}{\left\{}\newcommand{\rac}{\right\}}

\newcommand{\complex}{{\kern .1em {\raise .47ex
\hbox {$\scriptscriptstyle |$}}
    \kern -.4em {\rm C}}}
\newcommand{\real}{{{\rm I} \kern -.19em {\rm R}}}
\newcommand{\rational}{{\kern .1em {\raise .47ex
\hbox{$\scripscriptstyle |$}}
    \kern -.35em {\rm Q}}}
\renewcommand{\natural}{{\vrule height 1.6ex width
.05em depth 0ex \kern -.35em {\rm N}}}

\newcommand{\half}{\frac{1}{2}}
\newcommand{\pa}{\partial}

\newcommand{\dint}{\displaystyle{\int}}
\newcommand{\eg}{{\em e.g.,\ }}

\newcommand{\ie}{{{\em i.e.},\ }}

\newcommand{\twiddle}{\lower.9ex\rlap{$\kern -.1em\scriptstyle\sim$}}

\newcommand{\equ}[1]{(\ref{#1})}
\newcommand{\eq}{\begin{equation}}
\newcommand{\eqn}[1]{\label{#1}\end{equation}}
\newcommand{\eea}{\end{eqnarray}}
\newcommand{\eqa}{\begin{eqnarray}}
\newcommand{\eqan}[1]{\label{#1}\end{eqnarray}}
\newcommand{\ba}{\begin{array}}
\newcommand{\ea}{\end{array}}
\newcommand{\eqac}{\begin{equation}\begin{array}{rcl}}
\newcommand{\eqacn}[1]{\end{array}\label{#1}\end{equation}}

\makeatletter 
\@addtoreset{equation}{section}  
\makeatother 
 
\newcommand{\bz}{\begin{enumerate}}
\newcommand{\ez}{\end{enumerate}}

\newcommand{\ADS}{(A)dS}
\newcommand{\ads}{(a)ds}

\title{Five dimensional Chern-Simons Gravity \\
for the expanded (anti)-de Sitter gauge group C$_5$} 

\author{Matheus M.A. Paix\ao$^{1,\,2}$ and 
Olivier Piguet$^{1,\,}$\footnote{Present address: Pra\c ca Graccho Cardoso, 76/504, 45015-180 Aracaju, SE, Brazil}}

\begin{document}
\date{}    
\maketitle
\vspace{-5mm}

\begin{center}
\noindent
{\small $^1$Departamento de F\ii sica, Universidade Federal de 
Vi\c cosa (UFV)
Vi\c cosa, MG, Brazil.\\
$^2$Department of Astrophysics, Cosmology and Fundamental Interactions,
Brasilian Cen\-ter for Research in Physics,
CBPF, {BR-22290-180} Rio de Janeiro, Brazil.}


{\small\tt E-mails:  matheus.mapaixao@gmail.com, opiguet@yahoo.com }
\end{center}

\vspace{3mm}


\begin{abstract}
\quad  We study the Hamiltonian dynamics of a five-dimensional 
Chern-Simons theory for the gauge algebra $C_5$ of 
Izaurieta, Rodriguez and Salgado,  the so-called S$_H$-expans\ao\ of the 
5D (anti-)de Sitter algebra \textit{\ads}, based on the cyclic group
$\mathbb{Z}_4$. The theory consists 
 of a 1-form field containing the \textit{\ads}\ gravitation variables 
 and 1-form field transforming in the adjoint representation 
 of \textit{\ads}. The gravitational part of the action 
necessarily contains a term quadratic in the curvature, beyond 
the Einstein-Hilbert and cosmological terms, for any choice of 
 the  two independent coupling constants. 
The total action is also invariant under 
a new local symmetry, called ``crossed diffeomorphisms'',
 beyond the usual space-time diffeomorphisms. 
 The number of physical degrees of freedom is computed. 
 The theory is shown to be ``generic'' in the sense of 
 Ba\~nados, Garay and Henneaux, \ie the constraint 
 associated to the time diffeomorphisms 
is not independent from the other constraints.
\end{abstract}



\newpage

\tableofcontents

\section{Introduction}

In order to understand the universe  at the Planck scale, one needs 
a quantum theory which reduces to  General Relativity (GR) 
in the classical limit.
A  promising formalism is that of Loop Quantum Gravity 
(LQG)~\cite{Rovelli,Thiemann,Vidotto}. This approach uses 
the first-order formalism of GR, which is based on two pillars: 
Invariance under the local Lorentz transformations and invariance 
under the space-time diffeomorphisms.
 In this sense the Chern-Simons theories for gravitation 
 present an encouraging scenario. 
First, they are also 
background independent theories,  like GR, being invariant under 
the diffeomorphisms. Second, the actions are defined 
from invariant polynomials, which are 
gauge invariant by construction. Finally, they allow 
us to extend  naturally the local Lorentz invariance to 
a larger symmetry group including the  Poincar\'e  or the de Sitter 
or anti-de Sitter groups -- the latter being denoted by \ADS\ 
in this paper. The \ADS\ group is a "deformation" of the 
 Poincar\'e  group, the deformation parameter being the 
cosmological constant $\Lambda$, the special value 
$\Lambda=0$ corresponding to the  Poincar\'e  group.
 Pioneering works on the subject are those of Witten~\cite{Witten} 
for 3-dimensional space-time
and of Chamseddine~\cite{Chamseddine89,Chamseddine90}  
for space-times of dimension 5 or higher. More recent references together 
with a good review may be found in Hassaine and Zanelli's 
book~\cite{Zanelli}.

An analysis of the phenomenological aspects of 
5D Chern-Simons gravity models with dimensional reduction to 4D 
have been investigated, with results indicating their relevance 
as physical
theories~\cite{Aros,Salgado1',Tolosa-Zanelli,Vicosa1,Vicosa2,Bruno}. 

In the Hamiltonian formalism of Dirac~\cite{Dirac,Henneaux}, each local 
invariance is associated with a constraint which has to be 
solved at the quantum level. Unfortunately,  in gravitation theories, 
there are many difficulties with respect to the resolution 
of the Hamiltonian constraint,
the one corresponding to the invariance under the time 
diffeomorphisms~\cite{Rovelli,Thiemann,Vidotto}.
 In this sense, works by Ba\~nados, Garay and 
 Henneaux~\cite{Banados,Banados2} have made significant advances, 
showing the existence of so-called ``generic'' theories, 
where the constraint associated with the time diffeomorphisms 
is no longer independent, but can be seen as a combination of 
the constraints associated with gauge invariance and spatial
 diffeomorphisms. They have in particular shown that, 
 among others, the Chern-Simons theory in 5D space-time for the group 
\ADS$_5$ (\ie SO(1,5) or SO(2,4)) is in fact a generic theory, a result which
makes it an interesting candidate for a quantum gravity theory.

Another motivating factor for  the present work is found in a series 
of papers  on the  ``$S$-expansion'' of 
algebras~\cite{Salgado1,Salgado2,Salgado1',Salgado2'}. 
It is shown there that from any Lie algebra $\GG$ one can construct a 
new larger Lie algebra as the direct product 
$\GG_{\rm exp}$ = $\GG\times S$
of the starting Lie algebra  with a finite semi-group $S$. 
Thus, it is possible, \eg to obtain a group of symmetry wider 
than \ADS$_5$. This enables the introduction of new fields in the 
theory beyond the gravitation field. 
In particular, it was claimed 
in~\cite{Salgado1} that,   using an alternative expansion process 
called ``$H$-reduction'' leading to a symmetry algebra called $C_5$, 
it was possible 
to obtain a theory that reproduces exactly GR coupled 
with some matter fields, 
so that it would be a 
good candidate for the purpose of obtaining a quantum theory 
for gravitation  consistent with Einstein's 
theory\footnote{ See also~\cite{Concha} for an alternative approach.}.

The purpose of the present paper is, therefore, to construct the 
Chern-Simons action based on the  $S_H$-expanded algebra $C_5$
 of~\cite{Salgado1}.
Our main results are, first,  that the theory depends on 
two independent coupling constants\footnote{ The authors of~\cite{Concha}
arrive at the same conclusion, in a somewhat different interpretation frame. 
But our result enters in contradiction with~\cite{Salgado1}, where the theory
is claimed to depend on four independent coupling 
constants. The consequence of our result is to  
invalidate the $C_5$ symmetry of the Einstein-Hilbert model  
presented in Section 8 of~\cite{Salgado1}.}, 
second, that it is generic in the sense defined above, 
and third, that it is invariant under a new class of diffeomorphisms specific to these expanded algebra, 
which we call ``crossed diffeomorphisms''.

The paper is organized as follows. We make a brief review 
of \ads$_5$ Chern-Simons gravity in five dimensions in Section 2, 
 following essentially~\cite{Chamseddine89}.
The expansion process of Lie algebras  together with the  calculation of the 
$C_5$ invariant tensors are given in Section 3. 
In Section 4 we  check that the expansion considered by us actually leads 
to gravitational fields and ``fields of matter''. We will see 
that the pure gravitational part of the action
consists of an Einstein-Hilbert action term, a cosmological 
term and a term quadratic in the curvature, of the Gauss-Bonnet 
type. In Section 5 we  make a study of the dynamical structure 
of the theory, showing that it is also a generic theory.
The discussion of the generalized diffeomorphism invariance 
is presented in the final part of this section . The paper ends with our 
conclusions. Notations, conventions and some  technicalities
 can be found in Appendices  A and B.

The main results presented here constitute the content of a Master thesis 
defended by one of us~\cite{tese} at the Federal University of Vi\c cosa.
\section{Chern-Simons Gravity in 5D}\label{ads5}

 In this section we will present some results known in the literature 
on the Chern-Simons gravitation theories, important for 
the understanding of this work.  We follow Ref.~\cite{Chamseddine89}.
  The notations and conventions
 used are given in Appendix A.

\subsection{Chern-Simons Gravity in 5D for \ADS$_5$}\label{CS-ADS5}

 Chern-Simons theories occur only in odd dimensions. 
In this way we will start by treating the Chern-Simons 
theories in 5D,  whose gauge group  
is that of the transformations which leave invariant the metric 
$\eta_{MN}=\textrm{diag} \hspace{0.3em}(-1,1,1,1,1,s)$ of the 
internal space, where $M,N=0,1,\ldots,5$ and $s=\pm1$. 
For $s=+1$ we have the de Sitter group $SO(1,5)$ and for $s=-1$ 
we have the anti-de Sitter group $SO(2,4)$. 
For the moment we will not distinguish between them, 
simply calling them \ADS$_5$.

The associated Lie algebra,  denoted by \ads$_5$, 
consists of the $6\times6$ matrices $X$ whose elements have the form
$X_P{}^Q$ = $X_{PR}\eta^{RQ}$, with $X_{PR}=-X_{RP}$. 
A convenient basis is given by the 15 matrices $T^{MN}$ = $-T^{NM}$ 
defined by
\begin{equation}
(T_{MN})_P{}^Q=- \eta_{MP}\delta^Q_N +\eta_{NP}\delta^Q_M . 
\label{gerAdS}\end{equation}
The commutation relations are
\begin{equation}
\left[T_{MN},T_{PQ}\right]=T_{MP}\eta_{NQ}-T_{MQ}\eta_{NP}
-T_{NP}\eta_{MQ}+T_{NQ}\eta_{MP}, \label{comutacao}
\end{equation} 
which lead to the structure constants
\begin{equation}\ba{l}
f_{MN,PQ}{}^{RS} =  \frac{1}{2} \{ \eta_{MP}(\delta^R_N \delta^S_Q 
- \delta^S_N \delta^R_Q)+\eta_{NQ}(\delta^R_M \delta^S_P 
- \delta^S_M \delta^R_P)\es
+\eta_{PN}(\delta^R_Q \delta^S_M - \delta^S_Q \delta^R_i)
+\eta_{QM}(\delta^R_P \delta^S_N - \delta^S_P \delta^R_N)\}. \label{fAdS}
\ea\end{equation}

To construct the Chern-Simons action corresponding to \ADS$_5 $, 
we use the formalism of the differential forms. 
So, the fields are given by the $1$-form connection 
$A=A_{\mu}dx^{\mu}$, with $A_{\mu}=\dfrac{1}{2}A_{\mu}^{MN}T_{MN}$, 
where $\mu=0,1,\ldots,4$,  transforming under an 
infinitesimal gauge transformation as
\begin{equation}
\delta A= d\epsilon +[A,\epsilon],
\end{equation}
with $\epsilon$ = $\half\epsilon^{MN}T_{MN}$ an infinitesimal 
0-form Lie algebra 
valued  parameter. 
Said that, the Chern-Simons action, invariant up to boundary terms, is 
\eq\ba{l}
S= k\,\varepsilon_{MNPQRS}\dint \Lp A^{MN}\wedge dA^{PQ}\wedge dA^{RS}\wedge
+\frac{3}{2}A^{MN}\wedge(A^2)^{PQ}\wedge dA^{RS} \es
\phantom{S= \varepsilon_{MNPQRS}\dint \Lp }
+ \dfrac{3}{5}A^{MN}\wedge(A^2)^{PQ}\wedge(A^2)^{RS}\Rp,
\ea \eqn{CS5D} 
 where $\varepsilon_{MNPQRS}$, the 6D  completely antisymmetric
Levi-Civita tensor (with $\varepsilon_{012345}=1$), is an invariant 
rank 3 tensor\footnote{ Considering an antisymmetric pair such as 
$M,N$ as a single \ADS$_5$ index taking values from 1 to 15.} of \ADS$_5$.
The equations of motion derived from this action read
\begin{equation}
\varepsilon_{MNPQRS}F^{MN}\wedge F^{PQ}=0,
\label{FF=0}\end{equation} 
where $F^{MN}$ is the curvature $2$-form defined by
\eq 
F  = dA+A\wedge A = \half F^{MN} T_{MN},\quad
F^{MN}=dA^{MN}+A^M{}_U \wedge A^{UN}.
\eqn{curvature}
The solutions of \equ{FF=0} are not restricted  to the flat ones,
$F^{MN}=0$, as occurs in 
three-dimensional case~\cite{Witten}. 
Non-trivial solutions as well as cosmological models were studied 
in~\cite{Vicosa1,Vicosa2}.

In order to interpret this theory as a gravitation theory, 
one identifies the 15 generators $T_{MN}$ of the group \ADS$_5$ as 
the 10 generators $M_{AB}$ of the Lorentz group in 5D and 
the 5 generators $P_A$ of the generalized 
translations\footnote{The $P_A$ would be the translation generators in 
the Poincar\'e algebra case $s=0$.}, where $A,B=0,\ldots,4$:
\begin{equation}
M_{AB}=T_{AB}, \hspace{3.0em} P_A=\frac{1}{l}T_{A5}. \label{geradores}
\end{equation}
One writes accordingly the connection as 
\begin{equation}
A= \frac{1}{2}\omega^{AB}M_{AB}+ e^AP_A \label{decA}
\end{equation}
In \equ{geradores}, $l$ is a parameter with units of length 
(in the natural system of units),
necessary in order to take into account the difference between
 the dimensions of the vielbein $e^A$
and of the spin connection $\omega^{AB}$.
The commutation relations \eqref{comutacao} can be rewritten as
\begin{eqnarray}
\left[M_{AB},M_{CD}\right]
=M_{AC}\eta_{BD}-M_{AD}\eta_{BC}-M_{BC}\eta_{AD}+M_{BD}\eta_{AC},
\hspace{-3.1em}\nonumber\\
\left[M_{AB},P_{C}\right]=-P_{A}\eta_{BC}+P_{B}\eta_{AC},\hspace{8.0em}\\
\left[P_{A},P_{B}\right]=\frac{s}{l^2}M_{AB}, \hspace{12.7em} \nonumber
\end{eqnarray}
and the Chern-Simons action \eqref{CS5D} as\footnote{ From now on 
we skip the wedge symbol $\wedge$ for the external product of forms.}
\begin{equation}
S=k\int \varepsilon_{ABCDE}\left( e^AR^{BC}R^{DE} -\frac{2s}{3l^2}e^Ae^Be^CR^{DE} + \frac{1}{5l^4}e^Ae^Be^Ce^De^E \right), \label{S5}
\end{equation}
being 
\eq
R^{AB}=d\om^{AB}+\om^{AC} \om_C{}^B, 
\eqn{Riemann-curv} 
 the Riemann curvature $2$-form associated with the 
spin connection. The parameters $k$ and $l$ are related 
to the Newton's constant $G$ ($\propto s\, l^2/k$)
and to the cosmological constant $\Lambda$ ($\propto s/l^2$)~\cite{Vicosa2}.

As we can see we have the presence of the Einstein-Hilbert term 
and the cosmological constant one, in addition to the first term, 
which is of the Gauss-Bonnet type.

\subsection{Dynamics}\label{dynamics}

 The dynamics of Chern-Simons theories is best analysed 
via the Hamiltonian formalism of Dirac, identifying all the 
constraints of the theory and separating them into first and 
second class ones. An important concept to understand this dynamics 
in the context of Loop Quantum Gravity is that of ``generic theory'',
 first presented in~\cite{Banados}. 
 Let us use here the definition of genericity given in~\cite{Bruno}, 
 which although simpler than the first one, generalizes it. 
  Therefore, we will call a theory as generic if the constraints associated to the time diffeomorphisms is not an independent one, but can be expressed in terms of the other constraints that form a basis of all first-class constraints of the theory.

It has been shown in~\cite{Banados2} that the action  \eqref{CS5D}  leads to a theory with a total of $75$ constraints, 
$19$ of which are first class, of which 15 are associated with the \ADS\
gauge transformations and 4 with the spatial diffeomorphisms. 
The 56 remaining constraints are second-class. 
This leads to a theory with $13$ local degrees of freedom. 
Thus the theory is generic according to the definition given above. 
In particular, the Hamiltonian constraint, associated to the  time diffeomorphisms, is not independent,  
it can be expressed as a combination of the constraints 
associated with gauge transformations and spatial diffeomorphisms,
modulo field equations.

\section{The $C_5$ algebra}\label{C5-algebra}

\subsection{ $S$ and $S_H$ expansions}

 Following the construction of~\cite{Salgado2},
let us suppose   that we know a Lie algebra $\GG$ with basis $\{T_A\}$ and 
structure constants $f_{AB}{}^C$, 
the basic commutator being written as
\begin{equation}
	\left[T_A,T_B\right]=f_{AB}{}^CT_C.
\label{comm-rel}\end{equation}
Now, given a finite Abelian semi-group 
$S=\{\lambda_{\alpha};\alpha=1,\ldots,N \}$, 
it has been shown~\cite{Salgado2} that the direct 
product $S\times \GG$,  called the $S$-expansion of $\GG$,
with basis elements 
\eq 
T_{A\alpha}=\lambda_{\alpha}T_A
\eqn{exp-generators} 
and 
basic commutation rules defined by
\begin{equation}
[T_{A\alpha},T_{B\beta}]=[\lambda_{\alpha}T_{A},\lambda_{\beta}T_{B}] 
= \lambda_{\alpha}\lambda_{\beta}[T_{A},T_{B}],
\label{comSexp}\end{equation}
is another Lie algebra, of dimension equal to $N\, {\rm dim}(\GG$), 
called $S$-expanded algebra of $\GG$. So, we obtain a larger algebra 
having $\GG$ as a sub-algebra. 
Using the multiplication table of $S$, expressed by the 2-selector 
 $S_{\alpha\beta}{}^{\gamma}$:
\begin{equation}
S_{\alpha \beta}{}^{\gamma}
=\left\{\begin{array}{ll}1,\quad \lambda_{\alpha}\lambda_{\beta}
=\lambda_{\gamma}\\0,\quad  \textrm{otherwise},\end{array},\right.
\label{2-selector}\end{equation}
we may rewrite the commutation rules \equ{comSexp} as
\begin{equation}
[T_{A\alpha},T_{B\beta}]
=f_{A\alpha, B\beta}{}^{C\gamma}T_{C\gamma}, 
\label{comSexp'}\end{equation}
with the structure constants given by
\begin{equation}
f_{A\alpha, B\beta}{}^{C\gamma}=S_{\alpha\beta}{}^{\gamma}f_{AB}{}^C. 
\label{estruturaS}
\end{equation}
 The authors of~\cite{Salgado1} introduce another expansion, called 
$S_H$-expansion, consisting, in the case of $S$ being the 
cyclic group of even order  $\mathbb{Z}_{2n}$, in applying the conditions
\eq
T_{A,i}=\rho\,T_{A,i+n},\quad i=0,\cdots,n-1,
\eqn{H-cond}
with $\rho=-1$, on the generators $T_{A\a}$ ($\a=0,\cdots,2n-1$) 
of the $S$-expansion of $\GG$. They show that the resulting 
algebraic structure is a Lie algebra, denoted by $(\mathbb{Z}_{2n}\times\GG)_H$.
 We will use an alternative but equivalent approach.
We note that the conditions \equ{H-cond} are formally equivalent 
to  the conditions
\eq
\la_i=\rho\,\la_{i+n},\quad i=0,\cdots,n-1,
\eqn{H-cond-lambda}
on the elements of the group $S=\mathbb{Z}_{2n}$. 
Formally, the latter conditions amount 
to replace the multiplication table of $S=\mathbb{Z}_{2n}$ by the
$H$-reduced one, shown in
Table \ref{4-mult-table} for the special case  $n=2$ in which 
we will be interested in the following. Let us call
$\SSS$ this reduced group, which is Abelian.
\begin{table}[htb]
\begin{center}
	\begin{tabular}{l|llll}
		 & $\lambda_0$ & $\lambda_1$ & $\rho\lambda_0$ & $\rho\lambda_1$ \\
		\hline
		$\lambda_0$ & $\lambda_0$ & $\lambda_1$ & $\rho\lambda_0$ & $\rho\lambda_1$  \\
		$\lambda_1$ & $\lambda_1$ & $\rho\lambda_0$ & $\rho\lambda_1$ & $\lambda_0$  \\
		$\rho\lambda_0$ & $\rho\lambda_0$ & $\rho\lambda_1$ & $\lambda_0$ & $\lambda_1$  \\
		$\rho\lambda_1$ & $\rho\lambda_1$ & $\lambda_0$ & $\lambda_1$ & $\rho\lambda_0$
	\end{tabular}\end{center}\ \\[-12mm]
\caption{Multiplication table of the reduced group $\SSS$.}
\label{4-mult-table}
\end{table}	
Strictly speaking, the symbols "$\rho\la_i$" must
be considered as group elements independent of the
elements $\la_i$. It is only after substituting in the definition 
\equ{exp-generators} of the expanded generators, 
that $\rho$ will be considered as a number.

Let us consider the $S$-expanded algebra $\SSS\times$\ads$_5$,
with \ads$_5$ the (anti-)de Sitter algebra
defined in Subsection \ref{CS-ADS5} and  $\cal{S}$  
the Abelian group defined by the multiplication table shown in 
 Table \ref{4-mult-table}. From the 4 generators defined 
 by \equ{exp-generators}, two are independent, which may be taken as
 \eq 
T_{MNi}=\lambda_{i}T_{MN},\quad i=0,1,
\eqn{C5-exp-generators} 
with $T_{MN}$ given by  \eqref{gerAdS}.
Their commutations rules are obtained from \equ{comSexp},  
\equ{comSexp'}:
\begin{equation}
[T_{MNi},T_{PQj}]
=f_{MNi, PQj}{}^{RSk}T_{RSk}, 
\label{C5-comSexp'}\end{equation}
the structure constants being given by
\begin{equation}
f_{MNi, PQj}{}^{RSk}=S_{ij}{}^{k}f_{MN,PQ}{}^{RS}. 
\label{C5-estruturas}\end{equation}
with  $f_{MN,PQ}{}^{RS}$  given by \eqref{fAdS}. 
$S_{ij}{}^{k}$ ($i,j,k=0,1$) is the 2-selector (see definition \equ{2-selector})
corresponding to the mutiplication rules of
$\la_0$ and $\la_1$ given in the left upper quadrant 
of Table \ref{4-mult-table}. Its non-zero components 
are\footnote{We thank the authors of Ref.\cite{Salgado1} for pointing out to us
that this 2-selector, for $\rho=-1$, does not correspond to that of a semi-group, contrarily
to what we claimed in a previous version of this work.}
 \eq
S_{00}{}^0=1,\quad S_{01}{}^1=S_{10}{}^1=1,\quad S_{11}{}^0=\rho.
\eqn{2-selector-C5} 

We now  show that the Lie algebra 
we have constructed is identical, in the case where $\rho=-1$, to the Lie algebra 
$C_5$ = $(\mathbb{Z}_4\times$\ads$_5)_H$
of~\cite{Salgado1}. The structure constants of the latter are 
given by
$(\bar{S}_{i j}{}^{k}-\bar{S}_{i j}{}^{k+2})
f_{MN,PQ}{}^{RS}$,
with the indices $i,j,k$ taking the values 0, 1, and
 $\bar{S}_{\a\b}^\g$ ($\a,\b,\g=0,\cdots,3$)
is the 2-selector, as defined by Equ. \equ{2-selector}, for $\mathbb{Z}_4$. 
It is straightforward to check that they are equal to  our structure 
constants \equ{C5-estruturas}, which completes the proof.

 We observe that giving the value 1 to the factor $\rho$ leads to the 
$S$-expanded algebra $\mathbb{Z}_2\times$\ads$_5$, 
the selector \equ{2-selector-C5} being that of the cyclic group 
$\mathbb{Z}_2$.
We shall keep $\rho=\pm1$
unfixed,  treating both cases at the same time, 
but continuing to call this Lie algebra as $C_5$.
In both cases we have an algebra with twice as many independent 
generators (30) as for the \ads$_5$ algebra, hence the double 
of gauge fields, which may be divided in 15 gravitation and 
15 matter fields, as
will be made more precise later on in Section \ref{C5_action}.

\subsection{Invariant Tensors}

 As we mentioned, an important ingredient for constructing a 
Chern-Simons action invariant under a gauge group whose Lie 
algebra will be denoted by $\GG$, are the $\GG$-invariant 
tensors. Basically, a tensor in the adjoint representation like 
$g_{XYZ}$ is invariant if it obeys the relation\footnote{The indices 
$X, Y$, etc. can 
be multi-indices such as in Eqs. \equ{comSexp}, \equ{eq:invAds5} or \equ{invC5}.}
\begin{equation}
f_{TX}{}^{U}g_{U Y Z}+f_{T Y}{}^{U}g_{X U Z}+f_{T Z}{}^{U}g_{X Y U}=0. 
\label{eq:invg}
\end{equation}
where $f_{XY}{}^Z$ are the structure constants
 of $\GG$. We will restrict ourselves to rank 3 symmetric tensors.
Thus, the invariance condition of the \ads$_5$ case reads
\begin{equation}
f_{MN,M_1N_1}{}^{PQ} g_{PQ,M_2N_2,M_3N_3}+
f_{MN, M_2N_2}{}^{PQ}g_{M_1N_1,PQ, M_3N_3}+
f_{MN,M_3N_3}{}^{PQ}g_{M_1N_1,M_2N_2,PQ}=0 \label{eq:invAds5}
\end{equation}
From the structure constants \eqref{fAdS} and  the previous relation
one can show that $g_{MN,PQ,RS}$ = $\varepsilon_{MNPQRS}$, \ie 
the invariant tensor for the \ads$_5$ algebra is the Levi-Civita 
tensor of six indices.
Proceeding in an analogous way, the invariance condition 
\eqref {eq:invg} for the algebra $C_5 $ leads us to the 16 equations
\begin{eqnarray}
&&{S}_{i i_1}{}^{j}f_{MN,M_1N_1}{}^{PQ} g_{PQj, M_2N_2i_2,M_3N_3i_3}
+\nonumber\\ 
&& {S}_{i i_2}{}^{j} f_{MN,M_2N_2}{}^{PQ}g_{M_1N_1i_1, PQj,M_3N_3i_3}
+ \label{eq:invC5}\\
&&{S}_{i i_3}{}^{j}f_{MN,M_3N_3}{}^{PQ}g_{M_1N_1i_1 M_2N_2i_2,PQj}=0. \nonumber
\end{eqnarray}
with the 2-selector $S_{ij}{}^k$ 
 given in \equ{2-selector-C5}. 
  The indices $i,\,j,\cdots$ take the values 0, 1.
 The general solution of this system 
 for the $C_5$ invariant tensor, calculated in 
 App.  \ref{app-inv-tensors}, reads
\begin{eqnarray}
	g_{MNi, PQj,RSk}=
	c_{l}S_{i j k}{}^{l}\varepsilon_{MNPQRS}, \label{invC5}\end{eqnarray}
where the 3-selector~\cite{Salgado1} $S_{i j k}{}^{l}$ is  defined by
\begin{equation}
	S_{i j k}{}^{l}
=\left\{\begin{array}{lll}1,\  
 \lambda_{i}\lambda_{j} \lambda_{k}=\lambda_{l}\\
	\rho,\ \lambda_{i}\lambda_{j} \lambda_{k}= \rho \lambda_l\\
		0,\ \textrm{otherwise} \end{array}\right. 
		\label{3-selector}
\end{equation}
and  $c_0$ and $c_1$ are two arbitrary constants. 
So, the most general action  is an arbitrary linear combination 
of two invariant actions\footnote{The authors of 
reference~\cite{Salgado1}, who  consider the case where 
$\rho=-1$, give an invariant 
tensor with four independent parameters instead of two. 
We have checked that their solution satisfies the invariance 
condition \equ{eq:invC5} only 
if their four parameters obey 
two linear conditions and thus reduce to our solution \equ{invC5}.}. 

 We note that the expression \equ{invC5} 
is similar to that given by the Theorem 7.1 of [13],
with the difference that in our case, the selector $S_{ijk}{}^l$,
 like the 2-selector defined by \equ{2-selector-C5},
is not of a semi-group, but refers only  to
the first two elements of the group
defined by the multiplication Table \ref{4-mult-table}.

\section{Constructing the Chern-Simons action for  the $C_5$ 
algebra}\label{C5_action}

 The Chern-Simons action for the algebra $ C_5 $ is constructed in the same way 
as for the \ADS$_5$ case, changing only the symmetry  algebra. 
The gauge connection may be written as follows:
\begin{equation}
	\mathcal{A}=\frac{1}{2}\mathcal{A}^{MNi}T_{MNi}
	=\frac{1}{2}A^{MN}T_{MN0}+\frac{1}{2}B^{MN}T_{MN1},
\end{equation}
where we have separated  the  $\a=0$ and $\a=1$ components:
\begin{eqnarray}
	A^{MN}:=\mathcal{A}^{MN0} \quad
	B^{MN}:=\mathcal{A}^{MN1}.
\end{eqnarray}
It is also useful to write the $C_5$ connection $\AA$ in the form
\eq
\AA = \la_0 A + \la_1 B,\quad\mbox{with }
A=\frac{1}{2}A ^{MN}T_{MN},\ B=\frac{1}{2}B ^{MN}T_{MN},
\eqn{la-exp-AA}
which shows it explicitly as an $\SSS$-valued object.

The infinitesimal gauge transformations 
$\delta \mathcal{A} = d \mathcal{O} + [\mathcal{A},\mathcal{O}]$, 
of infinitesimal parameter
$\mathcal{O}=\lambda_0\omega+\lambda_1\eta$ read, for the fields $A$ and $B$:
\begin{eqnarray}
\delta A = \delta_{\omega}A+\delta_{\eta}A,\quad
\delta B =\delta_{\eta}B+\delta_{\omega}B,
\label{C5-gauge-tr}\end{eqnarray}
where
\eq\ba{ll}
\delta_{\omega}A=d\omega+[A,\omega], \quad&\delta_{\omega}B=[B,\omega],\es
\delta_{\eta}A=\rho[B,\eta], \quad& \delta_{\eta}B=d\eta+[A,\eta]. 
\ea\eqn{dw}
We observe that $\delta_{\omega}A$ is the transformation 
law of a connection for the $ (A) dS_5 $ group,  whereas 
the transformation $\d_\om B$ is that of a field in the 
adjoint representation of the same group. This allows us  
to identify the components of $A$ as the gravitation fields, 
whereas those of $B$ may be considered as the ``matter''  
fields of the theory. The transformations of parameter 
$\eta$ appear due to the process of expansion. 
Identifying the 5D Lorentz generators $M_{AB}$ = $T_{AB0}$ and 
the generalized translation generators $P_A$ = $T_{A50}/l$ 
(in the same way as in \equ{geradores}), 
we can explicitly identify the Lorentz connection forms 
$\om^{AB}$ and the 5-bein forms $e^A$ as the components 
of $A$ defined by Eq. \equ{decA}.

We can now decompose the $C_5$ curvature 
$\mathcal{F}=d\mathcal{A}+\mathcal{A}^2$ as
\eq
\mathcal{F}=\lambda_0 F+\lambda_1 G, 
\eqn{F}
where\footnote{ We use the multiplication table given by the upper left quadrant of Table \ref{4-mult-table} for $\ la_0$, $\la_1$.}
\eq 
F:=dA+A^2+\rho\, B^2,\quad G:=dB+[A,B]. 
\eqn{Fcom}
The next step is to calculate the Chern-Simons action for the 
expanded algebra $C_5$. In terms of the $C_5$ connection $\AA$, the action is  
\begin{eqnarray}
S=c_{l}S_{ijk}{}^{l}\varepsilon_{MNPQRS}\int \left(\mathcal{A}^{MNi}d\mathcal{A}^{PQj}d\mathcal{A}^{RSk} + \frac{3}{2}\mathcal{A}^{MNi}(\mathcal{A}^2)^{PQj}d\mathcal{A}^{RSk}+\hspace{0.8em}\right.\nonumber\\
\left.\frac{3}{5}\mathcal{A}^{MNi}(\mathcal{A}^2)^{PQj}(\mathcal{A}^2)^{RSk} \right). \label{CSC5}
\end{eqnarray}
The action contains two independent invariants, whose coefficients 
are the arbitrary coupling constants   $c_0$ and $c_1$. 
In terms of the $A$ and $B$ fields, the action takes the rather complicated form
\begin{eqnarray}
S=c_0\,\e_{MNPQRS}\int \left\{ \frac{{}}{{}}A^{MN}dA^{PQ}dA^{RS}+\rho\, A^{MN}dB^{PQ}dB^{RS}+2\rho\, B^{MN}dA^{PQ}dB^{RS}+ \right.\nonumber\\
\frac{3}{2}\left(A^{MN}(A^2)^{PQ}dA^{RS}+\rho\, A^{MN}(B^2)^{PQ}dA^{RS}+\rho\, A^{MN}[A,B]^{PQ}dB^{RS}\right.+\nonumber\\\left.\rho\, B^{MN}(A^2)^{PQ}dB^{RS}+B^{MN} (B^2)^{PQ}dB^{RS}+\rho\, B^{MN}[A,B]^{PQ}dA^{RS}\right)+\nonumber\\
\frac{3}{5}\left( {A}^{MN}(A^2)^{PQ}(A^2)^{RS}+{A}^{MN}(B^2)^{PQ}(B^2)^{RS}+2\rho\, {A}^{MN}(A^2)^{PQ}(B^2)^{RS}+\right.\nonumber\\\left.\rho\,{A}^{MN}[A,B]^{PQ}[A,B]^{RS}+2\rho\, {B}^{MN}[A,B]^{PQ}(A^2)^{RS}+2{B}^{MN} [A,B]^{PQ}(B^2)^{RS} \right) \left. \frac{{}}{{}}\right\}+\nonumber\\
c_1\,\varepsilon_{MNPQRS}\int \left\{ \frac{{}}{{}} B^{MN}dA^{PQ}dA^{RS}+\rho\, B^{MN}dB^{PQ}dB^{RS}+2A^{MN}dA^{PQ}dB^{RS}\right.+\nonumber\\
\frac{3}{2} \left( A^{MN}(A^2)^{PQ}dB^{RS}+\rho\, A^{MN}(B^2)^{PQ}dB^{RS}+ A^{MN}[A,B]^{PQ}dA^{RS}\right.+\nonumber\\\left. B^{MN}(A^2)^{PQ}dA^{RS}+\rho\, B^{MN} (B^2)^{PQ}dA^{RS}+\rho\, B^{MN}[A,B]^{PQ}dB^{RS}\right)\nonumber+\\
\frac{3}{5}\left( {B}^{MN}(A^2)^{PQ}(A^2)^{RS}+{B}^{MN}(B^2)^{PQ}(B^2)^{RS}+2\rho\, {B}^{MN}(A^2)^{PQ}(B^2)^{RS}+\right.\nonumber\\\left.\rho\,{B}^{MN}[A,B]^{PQ}[A,B]^{RS}+2{A}^{MN}[A,B]^{PQ}(A^2)^{RS}+2\rho\,{A}^{MN} [A,B]^{PQ}(B^2)^{RS} \right)\left.\frac{{}}{{}}\right\}. \label{Sexp}
\end{eqnarray}
If we put to zero the ``matter'' field $B$, we recover the action for
pure Chern-Simons gravitation in 5D given by Eq. \equ{CS5D}, or 
Eq. \equ{S5} in terms of the 5-bein $e^A$ and the Riemann curvature $R^{AB}$,
with $c_0$ as the coupling constant.
We see in particular that the Gauss-Bonnet term cannot be avoided, in
contradiction with the claim of~\cite{Salgado1}. 
 Let us mention however the possibility of achieving the separation of the 
Gauss-Bonnet term by assuming an alternative identification between the $C_5$
connection components $\AA^{MNi}$ and the gravity components
$\om^{AB}$, $e^A$ and the ``matter'' ones~\cite{Concha}. But in this case the
\ads\ invariance of the pure gravity part of the action does not hold any 
more.

\section{Dynamics}

 The dynamics of the $C_5$ theory is formally analogous to that of the  
 $(A)dS_5$  theory exposed in~\cite{Banados}. So, 
adapting the discussion of~\cite{Banados} to our present case, 
we will assume that the 5D 
manifold admits the topology $\mathbb{R}\times \Sigma$, where 
the real line $\mathbb{R}$ corresponds to time $x^0=t$, while  $\Sigma$ is
the dimension four space sheet with coordinates $x^a$, $a=1,\cdots,4$. 
Decomposing the $C_5$ connection according to this foliation,
\begin{equation}
	\mathcal{A}^{MNi}=\mathcal{A}^{MNi}_{0}dx^0
	+\mathcal{A}^{MNi}_adx^a,
\end{equation}
allows us to rewrite the Chern-Simons action in the form
\begin{equation}
S=\int\left( l^a_{MNi}\pa_t{\mathcal{A}}^{MNi}_a 
- \mathcal{A}^{MNi}_0K_{MNi} \right),
\end{equation}
where $K_{MNi}$ depends on the space components
of the $C_5$ curvature $\mathcal{F}$ and
$l_{MNi}^a$ on  the space components of the connection $\AA$ and
of the curvature $\mathcal{F}$: 
\eqa
K_{MNi}&=&-\frac{1}{32}c_{l}S_{ijk}{}^{l}
\varepsilon_{MNPQRS}\varepsilon^{abcd}\mathcal{F}^{PQj}_{ab}
\mathcal{F}^{RSk}_{cd}, \label{KC5}\\
l_{MNi}^a
&=&-\frac{1}{4}c_{l}S_{ijk}{}^{l}
\varepsilon_{MNPQRS}\varepsilon^{abcd}\mathcal{A}^{PQj}_{b}
\mathcal{F}^{RSk}_{cd}. \label{lC5}
\eea
The functional variations of the action with respect to $A_0^{MNi}$ 
and  $A_a^{MNi}$ yield the field equations
\eqa
K_{MNi}&=&0,\es
\Omega_{MNi, PQj}^{ab}(\pa_t\mathcal{A}^{PQj}_b
-D_b\mathcal{A}_0^{PQj})&=&0,
\eea
where
\begin{equation}
\Omega_{MNi, PQj}^{ab}
=-\frac{1}{2}c_{l}S_{ijk}{}^{l}\varepsilon_{MNPQRS}
\varepsilon^{abcd}\mathcal{F}^{RSk}_{cd} \label{WC5}.
\end{equation}

\subsection{Hamiltonian formalism  and constraints}

In order to pass from the Lagrangian to the Hamiltonian
 approach~\cite{Dirac,Henneaux}, we need the momenta conjugated to the 
 generalized coordinates, \ie to the fields 
 $\mathcal{A}_a^{MNi}$ and  $\mathcal{A}^{MNi}_0$: 
\eqa
p_{MNi}^a&=&l_{MNi}^a,\label{p-i}\\
p_{MNi}^0&=&0,\label{p-0}
\eea
 with $l_{MNi}^a$ given by \equ{lC5}.
These 150 relations between momenta and generalized coordinates are primary
constraints, which read
\eqa
\phi_{MNi}^a &:=& p_{MNi}^a-l_{MNi}^a\approx0, \label{phi-i}\\
\phi_{MNi}^0 &:=& p_{MNi}^0\approx0. \label{phi-0}
\eea
The Hamiltonian of the system is obtained by the usual Legendre transform
and the addition of the constraints multiplied by Legendre multiplier 
fields $\la$:
\begin{equation}
H= \int \left( \mathcal{A}^{MNi}_0 K_{MNi} 
+ \lambda_a^{MNi}\phi_{MNi}^a
+ \lambda_0^{MNi}\phi_{MNi}^0\right),
\end{equation}
 with $K_{MNi}$ given by \equ{KC5}.
The stability of the constraints \equ{phi-0} requires
$\dot{\phi}_{MNi}^0=\{\phi_{MNi}^0,H\}\approx0$, which 
implies 30 secondary constraints given by
\begin{equation}
K_{MNi}\approx 0. \label{phi-K}
\end{equation}
 The fields $\mathcal{A}^{MNi}_0$ play now the role of Lagrange multipliers, the constraints \equ{phi-0} becoming thus irrelevant.
So we are left with a total of 150 constraints in our theory. Those given by \eqref{phi-K} are conditions on the curvatures $ \mathcal{F} $. Thus not all the $ \mathcal{F}'s $ are independents. On the other end,
the constraints \eqref{phi-i} determine the $p_{MNi}^a$ in terms of the other variables.
Eqs. \equ{phi-K} for $i=0$ and $i=1$ read, in terms of the
fields $A$ and $B$ and with the definitions \equ{Fcom}:
\begin{eqnarray}
K_{MN0}=-\frac{1}{32}\varepsilon_{MNPQRS}\varepsilon^{abcd}\left[\left(F_{ab}^{PQ}F_{cd}^{RS}+\rho\, G_{ab}^{PQ}G_{cd}^{RS}\right)c_0+2F_{ab}^{PQ}G_{cd}^{RS}c_1\right]\approx 0, \label{K_MN0}\\
K_{MN1}=-\frac{1}{32}\varepsilon_{MNPQRS}\varepsilon^{abcd}\left[2\rho\, F_{ab}^{PQ}G_{cd}^{RS}c_0+\left(F_{ab}^{PQ}F_{cd}^{RS}+\rho\, G_{ab}^{PQ}G_{cd}^{RS}\right)c_1\right]\approx 0. \label{K_MN1}
\end{eqnarray}
However it will be opportune to substitute the constraint \equ{phi-K}
by the equivalent constraint
\begin{equation}
\mathcal{G}_{MNi}=-K_{MNi}+D_a\phi^a_{MNi}.
\label{constraintG}\end{equation}
Indeed, the Poisson parentheses of $ \mathcal{A} $ with $\mathcal {G}$,
\begin{equation}
\left\{\mathcal{A}_a^{MNi}(\vec{x}),\int \OO^{PQj}(\vec{y})\mathcal{G}_{PQj}\vec{(y)} \right\} = -D_a\OO^{MNi}(\vec{x}),
\end{equation} 
with $\OO^{MNi}$ an infinitesimal field in the $C_5$ algebra,
 show that the new constraints \equ{constraintG} generate the gauge transformations \equ{C5-gauge-tr}. 
The $\mathcal{G}_{MNi}$ and $\phi^a_{MNi}$ form a basis for the constraints and obey the (open) algebra relations
\begin{eqnarray}
\{ \mathcal{G}_{MNi}(\vec{x}),\mathcal{G}_{PQj}(\vec{y}) \} = S_{ij}{}^{k}f_{MN,PQ}{}^{RS}\mathcal{G}_{RSk}(\vec{x})\delta(\vec{x}-\vec{y}), \nonumber\es
\{ \phi^a_{MNi}(\vec{x}),\mathcal{G}_{PQj}(\vec{y}) \} = S_{ij}{}^{k}f_{MN,PQ}{}^{RS}\phi^a_{RSk}(\vec{x})\delta(\vec{x}-\vec{y}), \hspace{0.1em}\label{poisson}\es
\{ \phi^a_{MNi}(\vec{x}),\phi^b_{PQj}(\vec{y}) \} = \Omega^{ab}_{MNi,PQj}(\vec{x})\delta(\vec{x}-\vec{y}),\hspace{1.0em}\nonumber 
\end{eqnarray} 
where the $S_{ij}{}^{k}f_{MN,PQ}{}^{RS}$ are the 
$C_5$ structure constants and 
$\Omega^{ab}_{MNi,PQj}$ is given by \equ{WC5}. The constraints 
$\mathcal{G}_{MNi}$ appear here explicitly as first class 
ones\footnote{Since the Hamiltonian $H$ is a combination of 
constraints, we have that the Poisson brackets of the $\mathcal{G}$'s 
with it are also weakly zero.}, whereas
a quantity $N_2$  of the 120 $\phi^a_{MNi}$ will be second class, 
where $N_2$ is the rank of  $\Omega$ considered
as  a $120 \times 120$ matrix indexed by the multi-indices 
$\lac_{MNi}^a\rac$ and $\lac_{PQj}^b\rac$.
The theory does not allow for any further independent constraint. Indeed, 
the stability of the constraints $\mathcal{G}_{MNi}$ 
given by $\dot{\mathcal{G}}_{MNi}=0$ is automatically satisfied 
by the fact that they are first-class, 
while the stability of $\phi_{MNi}^a$ only carries us to 
restrictions on the Lagrange multipliers.

We turn now to the calculation of the rank $N_2$ of $\Omega$.
We already know the existence of four first class constraints, 
 given by the generators of the four spatial 
 diffeomorphisms~\cite{Banados}, which are 
 linear combinations of the  $\phi's $:
\eq
H_a=\mathcal{F}^{MNi}_{ab}\phi^b_{MNi}
=F_{ab}^{MN}\phi^b_{MN0}+G_{ab}^{MN}\phi^b_{MN1},
\eqn{diff-constr}
 which implies that $N_2\leq120-4=116$.
Inspired by the example of~\cite{Banados}, we take a  special case
of curvature with the following configuration:
\eq\ba{l}
F^{12} =G^{12}=dx^1dx^2+dx^3dx^4,\es
F^{34}=G^{34}=dx^1dx^2-dx^3dx^4,\es
F^{56}=G^{56}=dx^1dx^3+dx^2dx^4,\es
(\mbox{other components}=0),
\ea\eqn{FGAds5}
which can be derived from the potentials
\[\ba{l}
A^{12}=B^{12}=x^1dx^2+x^3dx^4, \es
A^{34}=B^{34}=x^1dx^2-x^3dx^4,\es
A^{56}=B^{56}=x^1dx^3+x^2dx^4,\es
(\mbox{other components}=0).
\ea\]
It is easy to verify that this configuration obeys the constraints 
\eqref{K_MN0} and \eqref{K_MN1}. Indeed, one sees that
$F^{12}F^{34}=G^{12}G^{34}=F^{12}G^{34}=G^{12}F^{34}=0$, 
as well $F^{12}F^{56}=G^{12}G^{56}=F^{12}G^{56}=G^{12}F^{56}=0$ 
and $F^{34}F^{56}=G^{34}G^{56}=F^{34}G^{56}=G^{34}F^{56}=0$, hence
$ K_{MN0}=0 $ and $K_{MN1}=0$.
We finally check that the corresponding matrix $\Omega$ has rank $N_2=112$.
 Besides this particular field configuration, we
 have analysed many numerical examples, all yielding a value 
$\leq 112$~\cite{tese}.
Thus we conclude, for the time being, that
\eq
112\leq N_2 \leq 116.
\eqn{rank112to116}
A more careful analysis of diffeomorphism invariance performed 
in the next subsection will show that this rank is in fact equal to 112.

\subsection{Crossed diffeomorphisms}

 We know that infinitesimal  diffeomorphisms 
$x'^\m=x^\m+\xi^\m(x)$
are given as Lie derivatives $\LL_\xi$ = $i_{\xi}d+di_{\xi}$, where $i_\xi$
is the contraction along the (infinitesimal) vector $\xi$.
Remembering that the $C_5$ connection is $S$-valued (see \equ{la-exp-AA}),
we consider an $S$-valued vector 
\eq
\xi=\lambda_{i}\xi^{i}=\lambda_0u+\lambda_1v,
\quad\mbox{with }\xi^0=u,\ \xi^1=v,
\eqn{S-valued-vector}
It is easy to see that the action as given by \equ{CSC5} 
in terms of the $C_5$ connection $\AA$ is invariant under the
generalized diffeomorphisms
defined as the Lie derivative along the $S$-valued vector 
\equ{S-valued-vector}:
\eq
\d_\xi\AA=\LL_\xi\AA.
\eqn{gen-diff}
Indeed, since $S=\int Q_5$ and $ Q_5 $ is a 5-form in five-dimensional 
space-time, its exterior derivative vanishes. Therefore
\begin{equation}
\delta_{\xi}S=\int\left(i_{\xi}dQ_5+di_{\xi}Q_5\right)=\int di_{\xi}Q_5
\end{equation}
is a boundary term.

From now on we specialize on the space generalized diffeomorphisms,
with $u=(u^a)$ and  $v=(v^a)$, $a=1,\cdots,4$.
Using $i_\xi=\la_0 i_u+\la_1 i_v$, we can write \equ{gen-diff}  for the 
components $A$ and $B$:
\begin{equation}
\delta A=\mathcal{L}_uA+\rho\, \mathcal{L}_vB, \quad 
\delta B=\mathcal{L}_uB+\mathcal{L}_vA,
\end{equation}
where $\rho=\pm1$ is the parameter entering in the 
multiplication laws given in Table \ref{4-mult-table}.

This way we have that
\begin{equation}
\delta_uA=\mathcal{L}_uA,\quad \delta_uB=\mathcal{L}_uB 
\label{dif_esp}\end{equation}
corresponds to the usual spatial diffeomorphisms, while the 
new symmetry transformations
\begin{equation}
\delta^{\times}_vA=\rho\,\mathcal{L}_vB,\quad 
\delta^{\times}_vB=\mathcal{L}_vA, 
\label{dif_cruz}\end{equation}
appearing as a result of the expansion process, will be
called  ``crossed diffeomorphisms''. Since $u$ and $v$ are 4-vectors, 
we have a total of 8 generalized diffeomorphisms.

These generalized diffeomorphisms obey commutation rules deduced
from the transformation law \equ{gen-diff} and
the Lie derivative commutator $[\LL_X,\LL_Y]=\LL_{[X,Y]}$ 
where $[X,Y]$ is the Lie bracket of the vectors $X$ and $Y$: 
\eq
[\delta_u,\delta_{u'}]=-\delta_{[u,u']},\quad
[\delta_u,\delta_{v}^{\times}]=-\delta_{[u,v]}^{\times},\quad
[\delta_v^{\times},\delta_{v'}^{\times}]=-\rho\,\delta_{[v,v']}.
\eqn{gen-diff-alg}

\subsection{Constraints associated with the crossed  dif\-fe\-omor\-ph\-isms}

Four first class constraints generating the usual spatial diffeomorphism 
are given by \equ{diff-constr}. We want now to find four first class 
constraints,
linear combinations of the primary constraints  of $\phi_{MNi}^a$, 
which generate the crossed  spatial diffeomorphims. 
 For this we will use the fact that the diffeomorphisms 
 \equ{gen-diff} given by the Lie derivative, can be equivalently represented by so-called improved diffeomorphisms~\cite{Banados}, given by
\eq
	\delta_{\xi}^{\rm impr}\mathcal{A}_{\mu}^{a}=\xi^{\nu}\mathcal{F}_{\nu\mu}^a,
\eqn{impr-diff}
since these differ from the Lie derivative only by a gauge transformation: 
\[
\xi^{\nu}\mathcal{F}_{\nu\mu}
=\mathcal{L}_{\xi}\mathcal{A}_\mu
-\delta_{(i\mathcal{A})}^{\rm gauge}\mathcal{A}_\mu.
\]
Recalling that $\AA$ as well as $\xi$ depend on the group  elements
$\la_0$ and $\la_1$ (see \equ{la-exp-AA} and \equ{S-valued-vector}), we can 
rewrite \equ{impr-diff} for the usual (parameter $u$) and
crossed (parameter $v$) diffeomorphisms of the fields $A$ and $B$, with
$F$ and $G$ their associated curvatures \equ{Fcom}
\eq\ba{ll}
\delta^{\rm impr}A=i_uF,\quad& \delta^{\rm impr}B=i_uG,\es
\delta^{\times,\rm impr}A=\rho\, i_vG,\quad& \delta^{\times,\rm impr}B=i_vF
\ea\eqn{u-v-impr}
The expression \equ{diff-constr} for the usual diffeomorphism constraint 
and a comparison between the two lines of \equ{u-v-impr} suggest 
the expression
\eq
H^\times_a
=\rho\, G_{ab}^{MN}\phi^b_{MN0}+F_{ab}^{MN}\phi^b_{MN1},
\eqn{cross-diff-constr}
for the  generators of the crossed  spatial diffeomorphism constraints. 
This is readily checked 
to be true:
\[\ba{c}
\left\{\mathcal{A}^{MNi}_a(x),
\int d^4x\, v^c(y)\left(\rho\, G_{cd}^{PQ}(y)\phi^d_{PQ0}(y)
+F_{cd}^{PQ}(y)\phi^d_{PQ1}(y)\right) \right\}\es
=v^c\left(\rho\, G_{ca}^{MN}+F_{ca}^{MN}\right)\hspace{0.3em}
	=i_v\lp \rho\, G^{MN}+F^{MN}\rp_a.
\ea\]
An important consequence of these considerations is that the 
$C_5$ theory is indeed generic according to the definition of genericity
given at the beginning of Section \ref{dynamics}.

\subsection{Counting the constraints and the degrees of freedom}

 Having thus found 8 first class constraints  
associated with the generalized diffeomorphisms, 
all of them being linear combinations of the 120 primary 
constraints $\phi^a_{MNi}$, we conclude that the rank $N_2$ 
of the matrix $\Omega$, hence the number of second class 
constraints, cannot exceed 112. 
In view of the inequality \equ{rank112to116}, we conclude that 
we have exactly $N_2=112$ second class constraints. Hence the number 
of first class constraints is $N_1=38$: the 30 constraints 
\equ{constraintG}, the 4 constraints \equ{diff-constr} and the 4 constraints 
\equ{cross-diff-constr} generating, respectively, 
the $C_5$ gauge transformations, the spatial diffeomorphisms 
and the  crossed spatial diffeomorphisms.

The number of physical degrees of freedom $N_{\rm d.o.f}$ is given by the 
formula~\cite{Banados,Banados2}
\eq
N_{\rm d.o.f}=\half\lp D_{\rm phase}-2N_1-N_2\rp,
\eqn{number-dof}
where $D_{\rm phase}=240$ is the dimension of the original phase
space of generalized coordinates and momenta $\AA^{MNi}_a$ and
$p_{MNi}^a$. The theory thus has 26 physical degrees of freedom.

\section{Conclusions}

Our first result concerns a fact about the algebra $C_5$ 
defined as the expansion of the 5D (anti-)de Sitter by 
the reduced group $\SSS$ whose multplication table is displayed in
Table \ref{4-mult-table}: we have shown  that  there are 
two independent symmetric invariant tensors of rank 3 in its 
adjoint representation, given by the equation \eqref{invC5}, 
instead of four as claimed by the authors of~\cite{Salgado1}, who
consider the same gauge group.  This result provides indeed a counter-example 
to the Theorem 2 in Section 5 of~\cite{Salgado1}.

The theory thus depends on two coupling constants, appearing
in the Chern-Simons action we have constructed for the 
$C_5$ gauge invariance. The $C_5$ connection 1-form $\AA$ decomposes into
the (anti-)de Sitter connection 1-form $A$ and a 1-form $B$ transforming 
in the adjoint representation of the latter algebra. 
The geometric part of the action, obtained by taking $B=0$, 
is identical to the \ads$_5$ one,  as originally given in~\cite{Chamseddine89}
and summarized in Section \ref{ads5}. When written in terms of the 
5-bein $e$ and the 5D Lorentz connection $\om$ (see \equ{S5}), the latter 
action shows, beyond the 
Einstein-Hilbert and cosmological terms a Gauss-Bonnet type 
term present which cannot be eliminated by any choice of 
the coupling constants -- in contradiction with the 
result of~\cite{Salgado1}.

 In third place, we have performed a complete canonical 
analysis of the dynamics of the theory, separating the total of
150 constraints into 112 second class ones and 38 first 
class ones, 30 of the latter being the generators of the 
$C_5$ gauge transformations and 8 being the generators 
of the generalized four-dimensional space diffeomorphisms. 
This result has allowed us to count the number of physical degrees of freedom
of the theory, 26. 

It is the canonical analysis which has  led us to  the 
extra local symmetry, called crossed 
diffeomeorphism invariance, which,  together with 
the usual  diffeomorphism  invariance, constitutes 
the generalized diffeomorphism invariance.
Another important by-product is that the theory is generic, \ie 
there is no independent constraint corresponding to the 
-- usual  and crossed -- temporal diffeomorphisms.

A possible unfolding of this work may be a realistic phenomenological 
analysis of the theory, making a dimensional reduction of 
5D to 4D, in a way similar to the one performed in~\cite{Vicosa1}, 
which presents solutions of the 
Schwarzschild type and others compatible with the $\Lambda$CDM 
cosmological model, despite the presence of the Gauss-Bonnet term. 
Another prospect
can be the quantization of the model. It would avoid the difficulty, found
in the quantization of General Relativity in $4D$, of 
solving the constraint associated with the invariance 
under the temporal diffeomorphisms, thanks to the genericity 
of the $C_5$ theory.

\subsection*{Acknowledgments}
 
 We thank the authors of Refs.~\cite{Salgado1} and~\cite{Concha}
 for enlighting   observations and discussions.\\
This work was partially funded by the 
Coordena\cao\ de Aperfei\c coa\-men\-to de Pessoal de N\ii vel 
Superior -- CAPES, Brazil (M.P.), by the 
 Conselho Nacional de Desenvolvimento Cien\-t\'{\i}\-fi\-co e
 Tecnol\'{o}gico -- CNPq, Brazil (M.P. and O.P.) and 
by the Funda\cao\ de Amparo a Pesquisa do Estado de 
Minas Gerais -- FAPEMIG, Brazil (O.P.). 

\appendix
\section*{Appendices}

\section{Notations and Conventions}

\begin{enumerate}
\item[$\bullet$] Units adopted are such that $c=1$.
\item[$\bullet$] The indices $M,N,P,Q,\ldots=0,1,\ldots,5$ are indices referring to \ADS$_5$.
\item[$\bullet$] The indices $A,B,C,D,\ldots=0,1,\ldots,4$ are Lorentz indices in 5D.
\item[$\bullet$]  The indices $\a,\b,\ldots=0,\cdots,3$ are indices referring to the group $\mathbb {Z}_4$.
\item[$\bullet$] The indices $i,j,\ldots=0,1$ are indices referring to the first two elements of the group $\mathbb {Z}_4$.
\item[$\bullet$] The indices $\mu,\nu,\ldots,\ldots=0,1,\ldots,4$ are space-time indices.
\item[$\bullet$] The indices $a,b,\ldots=1,2,3,4$ are spatial indices.
\item[$\bullet$] The metric for \ADS$_5$ is $\eta_{MN}=diag(-1,1,1,1,1,s)$, with $s=\pm 1$ ($+1$ refers to de Sitter and $-1$ to anti-de Sitter).
\end{enumerate}

\section{Invariant symmetric rank 3 tensors of $C_5$}\label{app-inv-tensors}

There is a unique invariant symmetric rank 3 tensor of the 
Lie algebra \ads$_5$,  solution of the conditions \equ{eq:invAds5}, given by
\eq
g_{MN,PQ,RS}=\varepsilon_{MNPQRS},
\eqn{ads-inv-tensor}
where $\varepsilon_{MNPQRS}$ is the 6D Levi-Civita antisymmetric tensor,
and each antisymmetric pair $MN$, etc. has to be considered as a 
multi-index, equivalent to an \ads$_5$ index taking values from 1 to 15.

In the case of the $C_5$ algebra, the invariance condition 
for the symmetric rank 3 invariant tensor $g_{MN{i_1},PQ{i_2},RS{i_3}}$
is given by 
\equ{eq:invC5}, where we have now multi-indices $MNi$ taking 30 values.
Since the indexes $i$, $i_1$, $i_2$, $i_3=0,1$ only, 
we have 16 possibilities of combining these indices, leading to 16 
equations that the invariant tensors must obey. Let us begin by the 
equation corresponding to
$(i,{i_1},{i_2},{i_3})$ = $(0,0,0,0)$:
\begin{eqnarray}
&&{S}_{0 0}{}^{0}f_{MN,M_1N_1}{}^{PQ} g_{PQ0, M_2N_20,M_3N_30}+\nonumber\\ 
&& {S}_{0 0}{}^{0} f_{MN,M_2N_2}{}^{PQ}g_{M_1N_10, PQ0,M_3N_30}+ \\
&&{S}_{0 0}{}^{0}f_{MN,M_3N_3}{}^{PQ}g_{M_1N_10 M_2N_20,PQ0}=0, \nonumber
\end{eqnarray}
Here, the $f_{\cdot\,\cdot,\,\cdot\,\cdot}{}^{\cdot\,\cdot}$ are 
the \ads$_5$ structure constants 
\equ{fAdS}.
The coefficients $S_{ij}{}^k$ being  here all 
equal to 1, we see that $g_{MN0,PQ0,RS0}$
obeys the \ads$_5$ invariance condition \eqref{eq:invAds5}. Hence
$g_{MN0,PQ0,RS0}=x_{000}\,\varepsilon_{MNPQRS}$, where $x_{000}$ 
is an arbitrary coefficient.
Proceeding in the same way for the seven other cases $(0,{i_1},{i_2},{i_3})$, 
we arrive at
\eq 
g_{MN{i_1},PQ{i_2},RS{i_3}} = x_{{i_1}{i_2}{i_3}} \e_{MNPQRS}.
\eqn{invC5-2}
The coefficients $x_{{i_1}{i_2}{i_3}}$ are completely symmetric 
in there indices due to the required symmetry of the tensor $g$. 
We have thus four 
independent parameters $x_{000}$, $x_{100}$, $x_{110}$ and $x_{111}$,
for the time being. Consider now the equations \equ{eq:invC5}
for $(i,{i_1},{i_2},{i_3})$ = $(1,1,0,0)$, using the result \equ{invC5-2}:
\[
\rho\, f_{MN,M_1N_1}{}^{PQ}x_{000} + 
(f_{MN,M_2N_2}{}^{PQ}+f_{MN,M_32N_3}{}^{PQ})x_{110}=0,
\]
where $\rho=\pm1$ is the value of the 2-selector $S_{11}{}^0$, 
see \equ{2-selector-C5}.
Due to the invariance condition for the Levi-Civita tensor $\e$
(see Eq. \equ{eq:invAds5} with $g_{.\,.\,.\,.\,.\,.}$ =
$\e_{.\,.\,.\,.\,.\,.}$), the latter equation reduces to
\[
f_{MN,M_1N_1}{}^{PQ}(\rho\, x_{000} -x_{110})=0,
\]
hence $x_{110}=\rho\, x_{000}$.
In the same way, now with  $(i,{i_1},{i_2},{i_3})$ = $(1,1,1,0)$, we find
$x_{111}=\rho\, x_{100}$.

Thus, the most general symmetric rank 3 tensor in the adjoint representation  of 
\ads$_5$ depends on two parameters, $c_0=x_{000}$ and $c_1=x_{100}$, so that
$x_{110}=\rho\, c_0$ and  $x_{111}=\rho\, c_1$.
With the 3-selector $S_{ijk}{}^{l}$ given by
 \eqref {3-selector}, we can now write 
$x_{i_1i_2i_3}$ = 
$c_{l}S_{i_1 i_2 i_3}{}^{l}$,
as can readily be verified.
This, with the use of the result \equ{invC5-2}, 
allows us to express the invariant tensor in the compact form 
\begin{eqnarray}
g_{MNi, PQj,RSk}=c_{l}S_{i j k}{}^{l}\varepsilon_{MNPQRS}. \nonumber
\end{eqnarray}


\end{document}